\documentclass[preprint,pre]{revtex4}
\usepackage{amsmath}
\usepackage{url}
\renewcommand{\vec}[1]{\boldsymbol{#1}}
\newcommand{\mat}[1]{\boldsymbol{\mathsf{#1}}}
\newcommand{\D}{\text{d}}
\newcommand{\chiab}{\chi_c}
\DeclareMathOperator{\diag}{diag}
\begin{document}

\title{Expressions for forces and torques in molecular simulations using rigid
  bodies}

\author{Michael P. Allen}
\affiliation{Department of Physics and Centre for Scientific Computing
\\
University of Warwick, Coventry CV4 7AL, United Kingdom
}
\author{Guido Germano}
\affiliation{Fachbereich Chemie, Philipps-Universit\"{a}t Marburg,
35032 Marburg, Germany}

\begin{abstract}
\noindent Expressions for intermolecular forces and torques, derived from pair
potentials between rigid non-spherical units, are presented. The aim is to give
compact and clear expressions, which are easily generalised, and which minimise
the risk of error in writing molecular dynamics simulation programs. It is
anticipated that these expressions will be useful in the simulation of liquid
crystalline systems, and in coarse-grained modelling of macromolecules.
\end{abstract}

\maketitle
\section{Introduction}
This paper is intended to clarify the way in which forces and torques may be
computed from an expression for the pair potential between two, generally
nonlinear, molecules.  These forces and torques are then typically incorporated
into molecular dynamics simulation programs
\cite{allen.mp:1989.j,sadus.rj:2002.a,rapaport.dc:2004.a}.  The particular
focus is on pair potentials which are expressed directly as functions of the
molecular centre-centre vector and orientational variables such as Euler angles,
quaternion (Euler) parameters, or rotation matrix elements.  Some examples,
based on potentials used in the simulation of liquid crystals and
macromolecules, are given, but the intention is also to provide a formalism that is
easily generalised.  The more common case of atomic site-site potentials is not
considered here. Several schemes exist for systematically ``coarse-graining''
such atomistically-detailed interaction potentials, often by performing
simulations over relatively short times and fitting either coarse-grained pair
distribution functions or aggregated forces
\cite{hahn.o:2001.a,kremer.k:2002.a,nielsen.so:2004.b,izvekov.s:2005.a,prapotnik.m:2005.a},
but this is also not the focus of the present paper. The aim is to streamline,
and make more robust, the process of analytically deriving forces and torques,
once the coarse-grained potentials have been determined, and parametrised in a
standard way.  Even with the help of symbolic algebra packages, performing this
task is prone to error and may result in molecular dynamics code which is
inefficient, hard to understand, and difficult to adapt. By the use of some
examples, based on spheroidal geometry, it is shown here that standard vector
and matrix notation helps to simplify the formulae, leading to an
improved insight into the machinery of the potentials as well as
giving a more reliable route to the simulation code.

Consider two nonlinear molecules A and B, position vectors
$\vec{r}_A$, $\vec{r}_B$.  Define the intermolecular vector
$\vec{r}=\vec{r}_{AB}=\vec{r}_A-\vec{r}_B$, distance $r=|\vec{r}|$,
and unit vector $\hat{\vec{r}}=\vec{r}/r$.  Specify the orientations
by the orthogonal rotation matrices $\hat{\mat{a}}$, $\hat{\mat{b}}$,
which convert from space-fixed $(xyz)$ to molecule-fixed $(123)$
coordinates \cite{goldstein.h:2002.a}:
\begin{equation*}
\hat{\mat{a}} = 
\begin{pmatrix} 
\hat{\vec{a}}_1\\ \hat{\vec{a}}_2\\ \hat{\vec{a}}_3
\end{pmatrix}
=
\begin{pmatrix} 
\hat{a}_{1x} & \hat{a}_{1y} &\hat{a}_{1z} \\
\hat{a}_{2x} & \hat{a}_{2y} &\hat{a}_{2z} \\
\hat{a}_{3x} & \hat{a}_{3y} &\hat{a}_{3z}
\end{pmatrix}
\;,
\qquad
\hat{\mat{b}} = 
\begin{pmatrix} 
\hat{\vec{b}}_1\\ \hat{\vec{b}}_2\\ \hat{\vec{b}}_3
\end{pmatrix}
=
\begin{pmatrix} 
\hat{b}_{1x} & \hat{b}_{1y} &\hat{b}_{1z} \\
\hat{b}_{2x} & \hat{b}_{2y} &\hat{b}_{2z} \\
\hat{b}_{3x} & \hat{b}_{3y} &\hat{b}_{3z}
\end{pmatrix}
\:.
\end{equation*}
The rows of these matrices are the three molecule-fixed orthonormal principal
axis vectors $\hat{\vec{a}}_m$, $\hat{\vec{b}}_m$, $m=1,2,3$.  Note that it is
possible to write $[\hat{\mat{a}}]_{m\mu} = [\hat{\vec{a}}_m]_\mu =
\hat{a}_{m\mu}$, $\mu=x,y,z$, without ambiguity.

It is assumed that the pair potential may be
written as a function 
$U = U(\vec{r},\hat{\mat{a}},\hat{\mat{b}})$
and that it is invariant to a global rotation of the coordinates.  Quite
generally the force $\vec{f}_A$ on A due to B will be the negative of the force
$\vec{f}_B$ on B due to A, so
\begin{equation*}
\vec{f}_A = -\vec{f}_B \equiv\vec{f}=-\frac{\partial U}{\partial\vec{r}} \:.
\end{equation*}
Depending on the form of the potential it may be possible to evaluate this
derivative directly, or it may be more convenient to separate the components
along, and perpendicular to, the line of centres:
\begin{equation}
\vec{f}  
=
-\frac{\partial U}{\partial r}
\frac{\partial r}{\partial\vec{r}}
-\frac{\partial U}{\partial\hat{\vec{r}}} \cdot
\left(\frac{\partial\hat{\vec{r}}}{\partial\vec{r}}\right)
=
-\frac{\partial U}{\partial r}\hat{\vec{r}}
- r^{-1}\frac{\partial U}{\partial\hat{\vec{r}}} \cdot
\bigl(\mat{1}-\hat{\vec{r}}\otimes\hat{\vec{r}}\bigr) \:.
\label{eqn:fABgen}
\end{equation}
On the right, $\otimes$ constructs a dyadic matrix from two vectors, and
$\mat{1}$ is the unit $3\times3$ matrix.
The torque calculation follows Price et al.\ \cite{price.sl:1984.a},
with minor variations.  Consider the negative of the derivative of $U$
with respect to rotation of molecule A through an angle $\psi$ about
any axis represented by a unit vector $\hat{\vec{\psi}}$. By
definition, this gives the corresponding component of the torque
$\vec{\tau}_A$, and the chain rule may be used to express this in
terms of the body-fixed unit vectors
\begin{equation*}
\hat{\vec{\psi}}\cdot\vec{\tau}_A 
= -\frac{\partial U}{\partial\psi}
= -\sum_{m} \frac{\partial U}{\partial\hat{\vec{a}}_m}
\cdot
\frac{\partial\hat{\vec{a}}_m}{\partial\psi} 
= -\sum_{m} \frac{\partial U}{\partial\hat{\vec{a}}_m}
\cdot\hat{\vec{\psi}}\times\hat{\vec{a}}_m
\:.
\end{equation*}
The last equation follows from the rotation formula
\cite{goldstein.h:2002.a} for any vector
$\partial\vec{e}/\partial\psi=\hat{\vec{\psi}}\times\vec{e}$.  Cyclic
permutation of the vectors in the scalar triple product then gives
\begin{equation*}
\hat{\vec{\psi}}\cdot\vec{\tau}_A = -\hat{\vec{\psi}}\cdot\sum_{m} \hat{\vec{a}}_m
\times\frac{\partial U}{\partial\hat{\vec{a}}_m} \:.
\end{equation*}
Choosing
$\hat{\vec{\psi}}$ to be each of the coordinate directions in turn allows the
identification of every component of the torque, and the expression for
$\vec{\tau}_B$ is similar
\begin{equation}
\label{eqn:tABgen}
\vec{\tau}_A = -\sum_{m} \hat{\vec{a}}_m
\times\frac{\partial U}{\partial\hat{\vec{a}}_m} \;,
\qquad
\vec{\tau}_B = -\sum_{m} \hat{\vec{b}}_m
\times\frac{\partial U}{\partial\hat{\vec{b}}_m} \:.
\end{equation}
In the succeeding sections, more specific examples of the above formulae will be
given for particular choices of pair potential $U$.
\section{Scalar Product Potentials}
A very common case, and a useful preamble for what follows, is when the pair
potential between A and B may be written in the form \cite{price.sl:1984.a}
\begin{equation}
U = U(\vec{r},\hat{\mat{a}},\hat{\mat{b}})
= U(r, \{\hat{\vec{a}}_m\cdot\hat{\vec{r}}\}, 
       \{\hat{\vec{b}}_n\cdot\hat{\vec{r}}\}, 
       \{\hat{\vec{a}}_m\cdot\hat{\vec{b}}_n\} )
\:.
\label{eqn:Uscalar}
\end{equation}
This is a function of the centre-centre separation $r$, and all possible scalar
products of the unit vectors $\hat{\vec{r}}$, $\hat{\vec{a}}_m$ and
$\hat{\vec{b}}_n$. It is manifestly invariant to a global rotation of
coordinates. The force may be written
\begin{align}
\vec{f} = -\frac{\partial U}{\partial\vec{r}}
&=
-\frac{\partial U}{\partial r}
\frac{\partial r}{\partial\vec{r}}
-\sum_{m} \left(
\frac{\partial U}{\partial(\hat{\vec{a}}_m\cdot\hat{\vec{r}})}
\frac{\partial(\hat{\vec{a}}_m\cdot\hat{\vec{r}})}{\partial\vec{r}}
+\frac{\partial U}{\partial(\hat{\vec{b}}_m\cdot\hat{\vec{r}})}
\frac{\partial(\hat{\vec{b}}_m\cdot\hat{\vec{r}})}{\partial\vec{r}}
\right)
\nonumber \\
&=
-\frac{\partial U}{\partial r} \hat{\vec{r}} 
-r^{-1}\sum_{m} \left(
\frac{\partial U}{\partial(\hat{\vec{a}}_m\cdot\hat{\vec{r}})}
\hat{\vec{a}}_m^\perp
+\frac{\partial U}{\partial(\hat{\vec{b}}_m\cdot\hat{\vec{r}})}
\hat{\vec{b}}_m^\perp
\right)
\:.
\label{eqn:fAB}
\end{align}
The sum ranges over all the orientation vectors on both molecules,
$\{\hat{\vec{a}}_m,\hat{\vec{b}}_m\}$. The notation
\begin{equation*}
\vec{v}^\perp \equiv
\vec{v}-(\vec{v}\cdot\hat{\vec{r}})\hat{\vec{r}}
= -\hat{\vec{r}} \times \bigl( \hat{\vec{r}} \times \vec{v} \bigr)
\end{equation*}
is short for the component of a vector $\vec{v}$ perpendicular to
$\hat{\vec{r}}$.  The derivatives of the potential appearing in
eqn~\eqref{eqn:fAB} are easily evaluated, assuming that it has the
general form of eqn~\eqref{eqn:Uscalar}.  The torque calculation once
more follows Price et al.\ \cite{price.sl:1984.a}.  The derivative of
$U$ with respect to rotation of molecule A through an angle $\psi$
about an axis $\hat{\vec{\psi}}$ takes the form
\begin{equation*}
\hat{\vec{\psi}}\cdot\vec{\tau}_A = -\frac{\partial U}{\partial\psi} = 
- \sum_{m}
 \frac{\partial U}{\partial(\hat{\vec{r}}\cdot\hat{\vec{a}}_m)}
\frac{\partial(\hat{\vec{r}}\cdot\hat{\vec{a}}_m)}{\partial\psi}
- \sum_{mn}
 \frac{\partial U}{\partial(\hat{\vec{b}}_n\cdot\hat{\vec{a}}_m)}
\frac{\partial(\hat{\vec{b}}_n\cdot\hat{\vec{a}}_m)}{\partial\psi}
 \:.
\end{equation*}
All combinations of unit vectors for which one, $\hat{\vec{a}}_m$, rotates with
the molecule while the other, $\hat{\vec{r}}$ or $\hat{\vec{b}}_n$, remains
stationary, appear on the right.  The rotation formula in this case gives
$\partial(\hat{\vec{v}}\cdot\hat{\vec{a}}_m)/\partial\psi =
\hat{\vec{v}}\cdot\hat{\vec{\psi}}\times\hat{\vec{a}}_m =
-\hat{\vec{\psi}}\cdot\hat{\vec{v}}\times\hat{\vec{a}}_m$ and once again
$\hat{\vec{\psi}}$ may be chosen arbitrarily. The explicit result for molecules
A and B is
\begin{subequations}
\label{eqn:tAB}
\begin{align}
\vec{\tau}_A &=
\sum_{m} 
\frac{\partial U}{\partial(\hat{\vec{a}}_m\cdot\hat{\vec{r}})}
\, \bigl(\hat{\vec{r}}\times\hat{\vec{a}}_m\bigr)
-
\sum_{mn} 
\frac{\partial U}{\partial(\hat{\vec{a}}_m\cdot\hat{\vec{b}}_n)}
\, \bigl(\hat{\vec{a}}_m\times\hat{\vec{b}}_n\bigr)
\;,
\label{eqn:tA}
\\
\vec{\tau}_B &=
\sum_{n} 
\frac{\partial U}{\partial(\hat{\vec{b}}_n\cdot\hat{\vec{r}})}
\, \bigl(\hat{\vec{r}}\times\hat{\vec{b}}_n\bigr)
+
\sum_{mn} 
\frac{\partial U}{\partial(\hat{\vec{a}}_m\cdot\hat{\vec{b}}_n)}
\, \bigl(\hat{\vec{a}}_m\times\hat{\vec{b}}_n\bigr)
\:.
\label{eqn:tB}
\end{align}
\end{subequations}
Note that the combination of eqns~\eqref{eqn:fAB}, \eqref{eqn:tA} and
\eqref{eqn:tB} gives 
\begin{equation}
\vec{\tau}_A+\vec{\tau}_B+\vec{r}\times\vec{f}=0
\:.
\end{equation}
This is the expression of local angular momentum conservation, which follows
directly from the invariance of the potential energy, eqn~\eqref{eqn:Uscalar}, to a
rotation of all the vectors $\hat{\vec{r}}$, $\hat{\vec{a}}_m$ and
$\hat{\vec{b}}_n$ together.
\section{Biaxial Gay-Berne, and Related, Potentials}
The original Gay-Berne family of potentials
\cite{corner.j:1948.a,berne.bj:1972.a,kushick.j:1976.a,gay.jg:1981.a} was
developed for molecules having the symmetry, and approximate shape, of uniaxial
ellipsoids.  They have now been generalised to include biaxiality
\cite{berardi.r:1995.a,cleaver.dj:1996.b,berardi.r:1998.a,berardi.r:2000.a}.
The general form is \cite{berardi.r:1998.a}
\begin{equation*}
U_\text{GB} = \varepsilon_1^\nu(\hat{\mat{a}},\hat{\mat{b}})
\, \varepsilon_2^\mu(\hat{\mat{a}},\hat{\mat{b}},\hat{\vec{r}})
\, U_\text{shape}\bigl(r,\sigma(\hat{\vec{r}},\hat{\mat{a}},\hat{\mat{b}})\bigr)
\end{equation*}
Here the two $\varepsilon$ terms represent orientation-dependent strength
factors, raised to powers $\mu$, $\nu$ which characterise different variants of
the model. The last term roughly defines the molecular shape, via an
orientation-dependent distance function $\sigma$ which will be discussed in more
detail below. $U_\text{shape}$ typically is based on a Lennard-Jones 12-6
potential
\begin{equation}
\label{eqn:ULJ}
 U_\text{shape}(r,\sigma) = 4\varepsilon_0 \, \left[
\left(\frac{\sigma}{r}\right)^{12} - \left(\frac{\sigma}{r}\right)^{6} \right]
\end{equation}
although, as shall be seen, this has been refined somewhat.  The
Gay-Berne potentials have been extensively used to model small organic
molecules \cite{gupta.s:1988.a,walsh.tr:2002.a,cacelli.i:2004.a} and
liquid crystalline systems
\cite{zannoni.c:2001.a,care.cm:2005.a}; %
linked Gay-Berne units may also be used in the coarse-grained
modelling of polymer chains including biological macromolecules
\cite{hahn.o:2001.a,ejtehadi.mr:2002.a}.

For illustrative purposes here, attention will focus on the ``shape'' term
above, and the energy-dependent $\varepsilon$ terms will be taken to be
constants. (Their contributions to forces and torques may be straightforwardly
included using similar methods). It is therefore assumed that the potential may
be written
\begin{equation}
\label{eqn:Ushape}
U = U\bigl(r,\sigma(\hat{\vec{r}},\hat{\mat{a}},\hat{\mat{b}})\bigr)
\:.
\end{equation}
The relation between the Gay-Berne distance function $\sigma$ and the geometry
of ellipsoids has been clarified by Perram et al.\
\cite{perram.jw:1984.a,perram.jw:1985.a,perram.jw:1996.a}. Define the matrices
\begin{equation}
\mat{A} 
= \sum_m a_m^2 \hat{\vec{a}}_m\otimes\hat{\vec{a}}_m
= \sum_m \vec{a}_m \otimes\vec{a}_m
\;,
\qquad
\mat{B} 
= \sum_m b_m^2 \hat{\vec{b}}_m\otimes\hat{\vec{b}}_m
= \sum_m \vec{b}_m\otimes\vec{b}_m
\end{equation}
(but note that our $\mat{A}$ is $\mat{A}^{-1}$ in
Ref.~\cite{perram.jw:1996.a}).  Here the $a_m$ and $b_m$ are semi-axis lengths
of the ellipsoidal particles, and these have been used to define the
un-normalised semi-axis vectors $\vec{a}_m=a_m\hat{\vec{a}}_m$ etc.  These
matrices may also be written in terms of the rotation matrices defined earlier
\cite{berardi.r:1995.a}
\begin{equation}
A_{\mu\nu} =  \sum_m a_m^2 \hat{a}_{m\mu}\hat{a}_{m\nu}
 = \sum_{mn} a_m a_n \delta_{mn} 
  \hat{a}_{m\mu}\hat{a}_{n\nu}
  = \sum_{mn} \bigl[\hat{\mat{a}}^\mathsf{T}\bigr]_{\mu m}
  a_m \delta_{mn} a_n \bigl[\hat{\mat{a}}\bigr]_{n\nu} \:.
\end{equation}
Define the diagonal shape matrix $\mat{S} = \diag(a_1,a_2,a_3)$, and the matrix
$\mat{a} = \mat{S}\hat{\mat{a}}$, whose rows are the un-normalised semi-axis
vectors $\vec{a}_m$, i.e.\ $\left[\mat{a}\right]_{m\mu}
=\left[\vec{a}_m\right]_\mu=a_{m\mu} = a_m \hat{a}_{m\mu}$, $m=1,2,3$. This
gives
\begin{equation}
\mat{A} = \hat{\mat{a}}^\mathsf{T} \mat{S}^2 \hat{\mat{a}}
 = \mat{a}^\mathsf{T} \mat{a} 
\:.
\end{equation}
Recall that the rotation matrix is orthogonal,
$\hat{\mat{a}}^\mathsf{T}\hat{\mat{a}} = \hat{\mat{a}}\hat{\mat{a}}^\mathsf{T} =
\mat{1}$, whereas this is not true for $\mat{a}$.

The diameter $\sigma$ is defined through the following equations
\cite{berardi.r:1995.a,cleaver.dj:1996.b,perram.jw:1996.a}
\begin{subequations}
\label{eqn:allsigma}
\begin{gather}
\label{eqn:sigma}
\sigma = 1/\sqrt{\varphi} \;, \\
\label{eqn:diam}
\varphi = \sigma^{-2} = \tfrac{1}{2} \thinspace
\hat{\vec{r}} \cdot \bigl(\mat{A}+\mat{B}\bigr)^{-1} \cdot \hat{\vec{r}}
\equiv
\tfrac{1}{2}\thinspace\hat{\vec{r}}\cdot\mat{H}^{-1}\cdot\hat{\vec{r}} \;,
\\ \text{where} \quad
\label{eqn:Hdef}
\mat{H} = \mat{A}+\mat{B} \:.
\end{gather}
\end{subequations}
All of the $\hat{\mat{a}}$ and $\hat{\mat{b}}$ dependence lies within the matrix
inverse $\mat{H}^{-1}$ while all the $\hat{\vec{r}}$ dependence lies outside it.
It is comparatively easy, though cumbersome, to evaluate the matrix inverse in a
suitable form, perform the scalar products with $\hat{\vec{r}}$ explicitly, and
use eqns~\eqref{eqn:fAB}, \eqref{eqn:tAB} for the forces and torques
\cite{berardi.r:1995.a,cleaver.dj:1996.b,allen.mp:1993.h}.  The following
alternative approach was suggested by Perram et al.\ \cite{perram.jw:1996.a}, and leads to
more compact expressions based on eqns~\eqref{eqn:fABgen}, \eqref{eqn:tABgen}.
\subsection{Scaled Potential}
\label{sec:scaled}
First consider the special class of potentials for which
\cite{perram.jw:1996.a}
\begin{gather}
\label{eqn:Uscaled}
U = U\bigl(r^2/\sigma^2(\hat{\vec{r}},\hat{\mat{a}},\hat{\mat{b}})\bigr)
 = U(\Phi)
\\
\text{where}\quad
\label{eqn:Phidef}
\Phi = r^2 \varphi = r^2/\sigma^2 = 
\tfrac{1}{2}\thinspace
\vec{r}\cdot\bigl(\mat{A}+\mat{B}\bigr)^{-1}\cdot\vec{r}
\equiv \tfrac{1}{2}\thinspace\vec{r}\cdot\mat{H}^{-1}\cdot\vec{r} \:.
\end{gather}
The simple form \eqref{eqn:ULJ} is an example. This form of potential
is simple to implement, but it does have the disadvantage that the
length scale is entirely determined by $\sigma$; hence, a
Lennard-Jones form will have an attractive range which is proportional
to the repulsive diameter at any given orientation of molecules, which
may be unphysical.  The shifted form of the potential (see
sec~\ref{sec:shifted}) was introduced by Gay and Berne
\cite{gay.jg:1981.a} to address this fault.  On the other hand, in
certain cases, e.g.\ for rapidly-varying, completely repulsive
potentials, this may not be so much a problem.

Following \cite{perram.jw:1996.a}, define an auxiliary vector $\vec{\kappa}$ by
solving the equations
\begin{equation}
\label{eqn:zetadef}
\mat{H}\cdot\vec{\kappa} = \vec{r} \qquad \Leftrightarrow \qquad
\vec{\kappa} = \mat{H}^{-1} \cdot \vec{r} \:.
\end{equation}
In what follows, there is no need to fully evaluate $\mat{H}^{-1}$.  In terms of
$\vec{\kappa}$
\begin{equation*}
\Phi = \tfrac{1}{2}\vec{r}\cdot\vec{\kappa}
\;,\qquad
\frac{\partial \Phi}{\partial\vec{r}} = 
\mat{H}^{-1}\cdot\vec{r}=\vec{\kappa} \:.
\end{equation*}
This last formula gives the interparticle force from 
\begin{equation}
\label{eqn:fscaled}
\vec{f} 
=
-\frac{\D U}{\D \Phi}
\frac{\partial \Phi}{\partial\vec{r}}
=
-\frac{\D U}{\D \Phi}\vec{\kappa} \:.
\end{equation}
To obtain the torque, write
\begin{equation}
\vec{\tau}_A 
= -\sum_{m} \hat{\vec{a}}_m\times
\frac{\partial U}{\partial\hat{\vec{a}}_m}
= -\sum_{m} \vec{a}_m\times
\frac{\partial U}{\partial\vec{a}_m}
=-\frac{\D U}{\D \Phi}\sum_{m} 
\vec{a}_{m}\times
\frac{\partial \Phi}{\partial\vec{a}_{m}} \:.
\label{eqn:tAscaled}
\end{equation}
The evaluation of $\partial\Phi/\partial\vec{a}_{m}$ proceeds as follows.
\begin{equation*}
\frac{\partial \Phi}{\partial a_{m\mu}} 
 = 
\tfrac{1}{2}\;\vec{r}\cdot
\frac{\partial\mat{H}^{-1}}{\partial a_{m\mu}}
\cdot\vec{r}
 = 
-\tfrac{1}{2}\;\vec{r}\cdot\mat{H}^{-1}
\frac{\partial \mat{H}}{\partial a_{m\mu}}
\mat{H}^{-1}\cdot\vec{r}
 = 
-\tfrac{1}{2}\;\vec{\kappa}\cdot
\frac{\partial \mat{A}}{\partial a_{m\mu}}
\cdot\vec{\kappa} \:.
\end{equation*}
The definition of $\mat{A}$ implies $\partial \mat{A}/\partial a_{m\mu} =
\hat{\vec{e}}_\mu \otimes\vec{a}_m + \vec{a}_m\otimes\hat{\vec{e}}_\mu$, where
$\hat{\vec{e}}_\mu$, $\mu=x,y,z$ is the appropriate unit vector in the
space-fixed Cartesian coordinate system.  Therefore 
\begin{equation*}
\partial \Phi/\partial
a_{m\mu} = - (\vec{\kappa}\cdot\vec{a}_m)\kappa_\mu
\qquad\Rightarrow\qquad
\partial \Phi/\partial \vec{a}_{m} = -
(\vec{\kappa}\cdot\vec{a}_m)\vec{\kappa} \:.
\end{equation*}
Combining with eqn~\eqref{eqn:tAscaled}, and applying a similar formula to
molecule B, gives the torques
\begin{subequations}
\begin{align}
\vec{\tau}_A &=\frac{\D U}{\D \Phi}\sum_m 
(\vec{\kappa}\cdot\vec{a}_m)(\vec{a}_{m}\times\vec{\kappa})
=
\frac{\D U}{\D \Phi}(\vec{\kappa}\cdot\mat{A}\times\vec{\kappa})
\;,
\\
\vec{\tau}_B &=
\frac{\D U}{\D \Phi}\sum_m
(\vec{\kappa}\cdot\vec{b}_m)(\vec{b}_{m}\times\vec{\kappa})
=
\frac{\D U}{\D \Phi}(\vec{\kappa}\cdot\mat{B}\times\vec{\kappa}) \:.
\end{align}
\end{subequations}
These equations are essentially those of Perram et al.\ \cite{perram.jw:1996.a}.  Using
eqn~\eqref{eqn:fscaled}, it is possible to write them in the form
\begin{subequations}
\label{eqn:tscaled}
\begin{alignat}{3}
\vec{\tau}_A 
&= -\vec{\kappa}\cdot\mat{A}\times\vec{f}
&&= \vec{r}_{CA}\times\vec{f}
&&= (\vec{r}_{C}-\vec{r}_{A})\times\vec{f} \;,
\\
\vec{\tau}_B 
&=-\vec{\kappa}\cdot\mat{B}\times\vec{f}
&&= \vec{r}_{CB}\times(-\vec{f})
&&=(\vec{r}_{C}-\vec{r}_{B})\times(-\vec{f})
\end{alignat}
\end{subequations}
where $\vec{r}_C$ is defined by
\begin{subequations}
\label{eqn:rcscaled}
\begin{align}
\vec{r}_C &= \vec{r}_A - \mat{A}\cdot\vec{\kappa} = \vec{r}_A + \vec{r}_{CA}
\\
&= \vec{r}_B + \mat{B}\cdot\vec{\kappa}= \vec{r}_B + \vec{r}_{CB}
\end{align}
\end{subequations}
(see Appendix~\ref{app:contact}).  The physical interpretation of this is that
the intermolecular force may be taken to act at the ``contact point''
$\vec{r}_C$, and the torques about the molecular centres are then given by the
usual moment formulae.  Conservation of angular momentum
follows immediately since
$\vec{\tau}_A+\vec{\tau}_B=(\vec{r}_{CA}-\vec{r}_{CB})\times\vec{f}
=-\vec{r}_{AB}\times\vec{f}=-\vec{r}\times\vec{f}$.

It is illuminating to consider the case of identical uniaxial particles.
Suppose the orientation of both particles is defined by unit vectors
$\hat{\vec{a}}_3$, $\hat{\vec{b}}_3$ along the 3-direction, and that the
semiaxes are given by $a_1=b_1=a_2=b_2 =\ell_{\perp}$, $a_3 = b_3 =
\ell_{\parallel}$.  Then, the orthonormality relation $\sum_m
\hat{\vec{a}}_m\otimes\hat{\vec{a}}_m = \mat{1}$ allows the orientation
matrices to be written in terms of $\hat{\vec{a}}_3$ and $\hat{\vec{b}}_3$ alone:
\begin{gather*}
\mat{A} = 
\ell_{\perp}^2 \mat{1}  
+ \bigl(\ell_{\parallel}^2-\ell_{\perp}^2\bigr)
\hat{\vec{a}}_3\otimes\hat{\vec{a}}_3
\;,
\qquad
\mat{B} = 
\ell_{\perp}^2 \mat{1}  
+ \bigl(\ell_{\parallel}^2-\ell_{\perp}^2\bigr)
\hat{\vec{b}}_3\otimes\hat{\vec{b}}_3
\;,
\\
\mat{H} = \mat{A}+\mat{B} = 2\ell_{\perp}^2 \mat{1} + 
\bigl(\ell_{\parallel}^2-\ell_{\perp}^2\bigr)\bigl(
\hat{\vec{a}}_3\otimes\hat{\vec{a}}_3+\hat{\vec{b}}_3\otimes\hat{\vec{b}}_3\bigr) \:.
\end{gather*}
Inverting such a matrix is a standard exercise. One route is to
observe that the orthonormal eigenvectors and eigenvalues are
\cite{cleaver.dj:1996.b}
\begin{align*}
\hat{\vec{e}}_\pm &=
\frac{\hat{\vec{a}}_3\pm\hat{\vec{b}}_3}{\sqrt{2(1\pm\hat{\vec{a}}_3\cdot\hat{\vec{b}}_3)}}
\;,
&
\lambda_\pm &= (\ell_\parallel^2+\ell_\perp^2) \pm 
(\ell_\parallel^2-\ell_\perp^2) \hat{\vec{a}}_3\cdot\hat{\vec{b}}_3 \;, \\
\hat{\vec{e}}_0 &= \hat{\vec{e}}_+ \times \hat{\vec{e}}_-
\;,
& \lambda_0 &= 2\ell_\perp^2 \;.
\end{align*}
Then both $\mat{H}$ and its inverse may be expressed as sums:
\begin{alignat*}{2}
\mat{H} &= \sum_m \lambda_m
\hat{\vec{e}}_m\otimes\hat{\vec{e}}_m
&& =  \lambda_0\mat{1}
+ (\lambda_+-\lambda_0) \hat{\vec{e}}_+\otimes\hat{\vec{e}}_+
+ (\lambda_--\lambda_0) \hat{\vec{e}}_-\otimes\hat{\vec{e}}_-
\;,
\\
\mat{H}^{-1} &= \sum_m \lambda_m^{-1}
\hat{\vec{e}}_m\otimes\hat{\vec{e}}_m
&& =  \lambda_0^{-1}\mat{1}
+ \bigl(\lambda_+^{-1}-\lambda_0^{-1}\bigr) \hat{\vec{e}}_+\otimes\hat{\vec{e}}_+
+ \bigl(\lambda_-^{-1}-\lambda_0^{-1}\bigr)
\hat{\vec{e}}_-\otimes\hat{\vec{e}}_-
\;,
\end{alignat*}
where the orthonormality property $\sum_m \hat{\vec{e}}_m\otimes\hat{\vec{e}}_m
= \mat{1}$ has been used. 

It is convenient to define the elongation parameter
$\chi=(\ell_\parallel^2-\ell_\perp^2)/(\ell_\parallel^2+\ell_\perp^2)$, so that
$1-\chi=2\ell_\perp^2/(\ell_\parallel^2+\ell_\perp^2)$; also scalar products of
the relevant unit vectors: $c_a=\hat{\vec{a}}_3\cdot\hat{\vec{r}}$,
$c_b=\hat{\vec{b}}_3\cdot\hat{\vec{r}}$,
$c_{ab}=\hat{\vec{a}}_3\cdot\hat{\vec{b}}_3$; and to set $\chiab=\chi c_{ab}$. Then
\begin{align*}
\mat{H}^{-1}  
&=\frac{1}{2\ell_\perp^2} \left\{
\mat{1} - \frac{\chi}{1-\chiab^2}
\left[\hat{\vec{a}}_3\otimes\hat{\vec{a}}_3+\hat{\vec{b}}_3\otimes\hat{\vec{b}}_3
- \chiab
(\hat{\vec{a}}_3\otimes\hat{\vec{b}}_3+\hat{\vec{b}}_3\otimes\hat{\vec{a}}_3)
\right]
\right\}
\\
&= \frac{1}{2\ell_\perp^2} \left\{
\mat{1} - \frac{\chi}{2}
\left[
\frac{(\hat{\vec{a}}_3+\hat{\vec{b}}_3)\otimes(\hat{\vec{a}}_3+\hat{\vec{b}}_3)}%
{1+\chiab }
+
\frac{(\hat{\vec{a}}_3-\hat{\vec{b}}_3)\otimes(\hat{\vec{a}}_3-\hat{\vec{b}}_3)}%
{1-\chiab }
\right]
\right\} \;,
\end{align*}
which gives the standard form \cite{berne.bj:1972.a} for identical uniaxial
particles
\begin{equation*}
\sigma^{-2}  
=\tfrac{1}{2}\thinspace\hat{\vec{r}}\cdot\mat{H}^{-1}\cdot\hat{\vec{r}}
= \sigma_0^{-2} \left\{
1 - \frac{\chi}{2}
\left[
\frac{(c_a+c_b)^2}%
{1+\chiab }
+
\frac{(c_a-c_b)^2}%
{1-\chiab }
\right]
\right\} \;,
\end{equation*}
with a width parameter $\sigma_0=2\ell_\perp$.  The auxiliary vector is
\begin{equation*}
\vec{\kappa} = \mat{H}^{-1} \cdot \vec{r}
=
\frac{r}{2\ell_\perp^2} \left\{
\hat{\vec{r}} - \frac{\chi}{1-\chiab^2}
\left[(c_a- \chiab c_b)\hat{\vec{a}}_3+(c_b- \chiab c_a)\hat{\vec{b}}_3
\right]
\right\} \:.
\end{equation*}
The position vector of the ``contact point'' $\vec{r}_C$ may be written
\begin{equation*}
\vec{r}_C=\tfrac{1}{2}(\vec{r}_A+\vec{r}_B)
 -\tfrac{1}{2} r \frac{\chi}{1-\chiab^2}
\left[(c_a- \chiab c_b)\hat{\vec{a}}_3-(c_b- \chiab c_a)\hat{\vec{b}}_3
\right] \:.
\end{equation*}
Having chosen the form of $U(\Phi)$, these last two equations make it easy to
calculate forces and torques through eqns~\eqref{eqn:fscaled} and
\eqref{eqn:tscaled}. 
\subsection{Shifted Potential}
\label{sec:shifted}
The shape-dependent part of the Gay-Berne potential \cite{gay.jg:1981.a} does
not have the simple form of eqn~\eqref{eqn:Uscaled}, but uses the more general
eqn~\eqref{eqn:Ushape}, with the definition of $\sigma$ given in
eqn~\eqref{eqn:diam}. This means that the potential is \emph{not} a function of
the single variable $\Phi=r^2/\sigma^2$, but instead depends on both $r$ and
$\sigma$ (or $\varphi=\sigma^{-2}$) separately,
typically through the shifted form
\begin{equation}
\label{eqn:rhodef}
\varrho = \frac{r-\sigma+\sigma_\text{min}}{\sigma_\text{min}} \;,
\end{equation}
where $\sigma_\text{min}$ is a constant.  Define as before
$\vec{\kappa}=\mat{H}^{-1}\cdot\vec{r}$, $\varphi =
\tfrac{1}{2}\hat{\vec{r}}\cdot\mat{H}^{-1}\cdot\hat{\vec{r}} = \tfrac{1}{2}
r^{-1}\hat{\vec{r}}\cdot\vec{\kappa}$, so
\begin{equation*}
\frac{\partial \varphi}{\partial\hat{\vec{r}}} = 
\mat{H}^{-1}\cdot\hat{\vec{r}}
=
r^{-1}\vec{\kappa} \:.
\end{equation*}
This gives the interparticle force
\begin{equation}
\vec{f} 
=
-\frac{\partial U}{\partial r}\hat{\vec{r}}
-r^{-1}\frac{\partial U}{\partial \varphi}
\frac{\partial \varphi}{\partial\hat{\vec{r}}} \cdot
\bigl(\mat{1}-\hat{\vec{r}}\otimes\hat{\vec{r}}\bigr)
=
-\frac{\partial U}{\partial r}\hat{\vec{r}}
-r^{-2}\frac{\partial U}{\partial \varphi}
\bigl[\vec{\kappa}-(\vec{\kappa}\cdot\hat{\vec{r}})\hat{\vec{r}}\bigr] \:.
\label{eqn:shff}
\end{equation}
The torque derivation proceeds exactly as before, and the results are
\begin{equation}
\vec{\tau}_A 
=
r^{-2}\frac{\partial U}{\partial \varphi}
(\vec{\kappa}\cdot\mat{A}\times\vec{\kappa})
\;, \qquad
\vec{\tau}_B 
=
r^{-2}\frac{\partial U}{\partial \varphi}
(\vec{\kappa}\cdot\mat{B}\times\vec{\kappa}) \:.
\end{equation}
Identifying part of the force $\vec{f}_\kappa = -r^{-2}(\partial
U/\partial\varphi)\vec{\kappa}$ these equations may be written
\begin{equation}
\label{eqn:shft}
\vec{\tau}_A 
= -\vec{\kappa}\cdot\mat{A}\times\vec{f}_\kappa
= \vec{r}_{CA}\times\vec{f}_\kappa \;, \qquad
\vec{\tau}_B 
=-\vec{\kappa}\cdot\mat{B}\times\vec{f}_\kappa
= \vec{r}_{CB}\times(-\vec{f}_\kappa) \;,
\end{equation}
where $\vec{r}_{C}$ is given by eqn~\eqref{eqn:rcscaled} as before. 
Angular momentum conservation
follows directly, since
$\vec{f}-\vec{f}_\kappa$ is parallel to $\hat{\vec{r}}$, so
$\vec{r}\times\vec{f}=\vec{r}\times\vec{f}_\kappa$.

As an example, consider the repulsive potential
\begin{equation}
\label{eqn:ULJcut}
U = \begin{cases}
4 \varepsilon_0\left(\varrho^{-12}-\varrho^{-6} \right) + 
 \varepsilon_0 \;, & \varrho^6 < 2 \\
0 \;, & \varrho^6 > 2
\end{cases} 
\end{equation}
with $\varrho$ given by eqn~\eqref{eqn:rhodef} and $\sigma$ given by
eqn~\eqref{eqn:diam}.  It is sensible to set $\sigma_{\text{min}}$ to
be the minimum value of $\sigma$ over all relative orientations. The
potential then takes a constant value $U=+\varepsilon_0$ when
$r\rightarrow \sigma$, for any orientation, and diverges as
$r\rightarrow \sigma-\sigma_\text{min}$.  For example, for the
identical uniaxial particles of the previous section,
$\sigma_\text{min}=\sigma_0=2\ell_\perp$ for the prolate case
$\ell_\parallel>\ell_\perp$, corresponding to side-by-side
arrangements of the particles; but $\sigma_\text{min}=2\ell_\parallel$
for oblate, disk-like, shapes, corresponding to the face-to-face
orientation \cite{bates.ma:1996.a}.  In the general case,
eqn~\eqref{eqn:diam} shows that if the minimum particle dimensions are
$a_{\text{min}}=\min(a_1,a_2,a_3)$,
$b_{\text{min}}=\min(b_1,b_2,b_3)$, then
\begin{equation*}
\sigma_{\text{min}}^2 = 2H_{\text{min}} =2(a_{\text{min}}^2 + b_{\text{min}}^2)
\:.
\end{equation*}
The geometrically correct value for spheroids would be $\sigma_{\text{min}} =
a_{\text{min}} + b_{\text{min}}$ but this distance function does not faithfully
represent the geometry of spheroids \cite{perram.jw:1996.a}.  Then
\begin{equation*}
-\frac{\partial U}{\partial\varphi} =
-\frac{\partial U}{\partial\varrho}\frac{\partial\varrho}{\partial\sigma}
\frac{\D\sigma}{\D\varphi}
=
24\varepsilon_0\left(2\varrho^{-13}-\varrho^{-7}\right)
\sigma^3/2\sigma_{\text{min}}
\:.
\end{equation*}
Also
\begin{equation*}
-\frac{\partial U}{\partial r} =
-\frac{\partial U}{\partial\varrho}\frac{\partial\varrho}{\partial r}
=
24\varepsilon_0\left(2\varrho^{-13}-\varrho^{-7}\right)/\sigma_{\text{min}}
\:.
\end{equation*}
From these two expressions, together with
eqns~\eqref{eqn:shff}--\eqref{eqn:shft}, the forces and torques may be easily
calculated.  

The above equations generalise simply to the case of the full
Gay-Berne potential, where the orientation-dependent energy functions
$\varepsilon_1$ and $\varepsilon_2$ are included
\cite{berardi.r:1998.a}. When the Lennard-Jones form
\eqref{eqn:ULJcut} is not truncated, the limiting long-range form is
proportional to $r^{-6}$, as might be expected. However, a well-known
criticism of the potential is that the full orientation-dependence is
preserved at long range, whereas the van der Waals attractions in real
molecules become isotropic at long range. It should be noted that
Everaers and Ejtehadi \cite{everaers.r:2003.a} have derived a
coarse-grained biaxial potential, called the RE-squared model, using
Hamaker theory from colloid science. The result has features in common
with the biaxial Gay-Berne potential, but also has the correct
long-distance form.  Babadi et al.\ \cite{babadi.m:2006.a} have
shown how to parametrise the RE-squared model using atomistic
representations of small molecules. Forces and torques may be derived
from this potential in a way analogous to that described above.
\section{Gaussian Potential}
\label{sec:gaussian}
Some years ago, Smith and Singer \cite{smith.w:1985.a} proposed an
extension of the approach of Berne and Pechukas
\cite{berne.bj:1972.a}, which they incorporated into the CCP5 Library
program \textsc{mdzoid}.  Each molecule is represented by a normalised
Gaussian distribution of particle density, centred at $\vec{r}_A$ and
$\vec{r}_B$, characterised by matrices $\mat{A}$ and $\mat{B}$:
\begin{subequations}
\label{eqn:rhoABdef}
\begin{align}
\rho_A(\vec{r}) &= \frac{1}{\sqrt{8\pi^3 |\mat{A}|}}
~\exp\bigl[-\tfrac{1}{2}
(\vec{r}-\vec{r}_A)\cdot\mat{A}^{-1}\cdot(\vec{r}-\vec{r}_A)\bigr]
\;,
\\
\rho_B(\vec{r}) &= \frac{1}{\sqrt{8\pi^3 |\mat{B}|}}
~\exp\bigl[-\tfrac{1}{2}
(\vec{r}-\vec{r}_B)\cdot\mat{B}^{-1}\cdot(\vec{r}-\vec{r}_B)\bigr]
\:,
\end{align}
\end{subequations}
where $|\cdots|$ is the determinant of a matrix.  The interaction
between volume elements $\D\vec{r}_a$, $\D\vec{r}_b$ at $\vec{r}_a$
and $\vec{r}_b$, respectively, is represented as a Gaussian function
of their separation
\begin{equation}
\label{eqn:sspot}
u(\vec{r}_a-\vec{r}_b)\D\vec{r}_a\D\vec{r}_b 
=C\ell^{-3}\exp\bigl(-\tfrac{1}{2}\left|\vec{r}_a-\vec{r}_b\right|^2/\ell^2\bigr)~
\D\vec{r}_a\D\vec{r}_b \;,
\end{equation}
characterised by a range parameter $\ell$ and a strength parameter $C$
(with units of energy multiplied by volume).  (More generally, Smith
and Singer \cite{smith.w:1985.a} suggested fitting a desired
atom-atom potential with a sum of such Gaussians, with positive or
negative coefficients).  The potential energy between molecules A and
B is obtained by integrating $u(\vec{r}_a-\vec{r}_b)$ over the density
distributions $\rho_A(\vec{r}_a)$ and $\rho_B(\vec{r}_b)$; the
integrals may be performed analytically (see Appendix
\ref{app:gausspot}) and the result is, with $\vec{r}=\vec{r}_{AB}$,
\begin{subequations}
\label{eqn:gausspot}
\begin{align}
U &=
C|\mat{H}|^{-1/2} \exp(-\Phi) \;,
\\
\text{with}\quad \mat{H} &= \ell^2\mat{1}+\mat{A}+\mat{B} \;,
\\
\text{and}\quad\Phi &=\tfrac{1}{2}\vec{r}\cdot
\bigl(\ell^2\mat{1}+\mat{A}+\mat{B}\bigr)^{-1}
\cdot\vec{r}
=\tfrac{1}{2}\vec{r}\cdot\mat{H}^{-1}\cdot\vec{r}
\:.
\end{align}
\end{subequations}
In the limit $\ell\rightarrow0$, $\mat{H}$ and $\Phi$ reduce to the definitions
\eqref{eqn:Hdef}, \eqref{eqn:Phidef} seen before.  This potential belongs to the
class of ``core-softened'' potentials, which have recently attracted attention
in condensed matter physics.  The interaction between polymer coils at low
concentration, for instance, is well represented by a (spherically symmetric)
Gaussian potential \cite{krakoviack.v:2003.a}.

The forces are derived exactly as for the scaled potential,
sec.~\ref{sec:scaled}: 
\begin{equation}
\label{eqn:fgaussian}
\vec{f} 
=
-\frac{\partial U}{\partial\Phi}
\frac{\partial \Phi}{\partial\vec{r}}
=
-\frac{\partial U}{\partial\Phi}\mat{H}^{-1}\cdot\vec{r}
=
U\mat{H}^{-1}\cdot\vec{r}
=
U\vec{\kappa} \;,
\end{equation}
with $\vec{\kappa}=\mat{H}^{-1}\cdot\vec{r}$ as before.  The torques
have contributions from the dependence of $U$ on both $\Phi$ and
$|\mat{H}|$.  The $\Phi$ contributions are given by
\begin{alignat*}{3}
\vec{\tau}_A^\Phi
&= -\vec{\kappa}\cdot\mat{A}\times\vec{f}
&&= \vec{q}_{A}\times\vec{f} \;,
\\
\vec{\tau}_B^\Phi 
&=-\vec{\kappa}\cdot\mat{B}\times\vec{f}
&&= \vec{q}_{B}\times(-\vec{f}) \;,
\end{alignat*}
which are eqns~\eqref{eqn:tscaled}, with $\vec{r}_{CA}\rightarrow
\vec{q}_{A}=-\mat{A}\cdot\vec{\kappa}$, $\vec{r}_{CB}\rightarrow
\vec{q}_{B}=\mat{B}\cdot\vec{\kappa}$.  The reason for the change in notation is
that these equations do not define a contact point $\vec{r}_{C}$ as before in
eqns~\eqref{eqn:rcscaled}, because the new definition of $\mat{H}$ gives
\begin{equation}
\vec{r} + \vec{q}_{A} - \vec{q}_{B} = \ell^2 \vec{\kappa} \;,
\label{eqn:newrcscaled}
\end{equation}
rather than zero.  Using the expression
$\frac{1}{2}|\mat{H}|^{-1}(\partial|\mat{H}|/\partial \vec{a}_{m}) =
\mat{H}^{-1}\cdot\vec{a}_{m}$ (see appendix \ref{app:gausspot}), the
$|\mat{H}|$-dependent terms in $ \vec{\tau}_A$ give
\begin{equation}
\label{eqn:mpa1}
-\frac{\partial U}{\partial|\mat{H}|}\sum_{m} 
\vec{a}_{m}\times
\frac{\partial |\mat{H}|}{\partial\vec{a}_{m}}
=\tfrac{1}{2}|\mat{H}|^{-1}U\sum_{m} 
\vec{a}_{m}\times
\frac{\partial |\mat{H}|}{\partial\vec{a}_{m}}
=U\sum_{m} 
\vec{a}_{m}\times\mat{H}^{-1}\cdot\vec{a}_{m} \;.
\end{equation}
With the use of the Levi-Civita tensor $\mat{\epsilon}$ the final results may be
expressed 
\begin{subequations}
\begin{alignat}{2}
\vec{\tau}_A
&=U\sum_{m} \vec{a}_{m}\times\mat{H}^{-1}\cdot\vec{a}_{m}
+\vec{q}_{A}\times\vec{f}
&&=U\mat{\epsilon}:\mat{H}^{-1}\mat{A}+\vec{q}_{A}\times\vec{f} \;,
\label{eqn:tAzoid}
\\
\vec{\tau}_B 
&=U\sum_{m} \vec{b}_{m}\times\mat{H}^{-1}\cdot\vec{b}_{m}
-\vec{q}_{B}\times\vec{f}
&&=U\mat{\epsilon}:\mat{H}^{-1}\mat{B}+\vec{q}_{B}\times(-\vec{f}) \:.
\label{eqn:tBzoid}
\end{alignat}
\end{subequations}
Smith and Singer \cite{smith.w:1985.a} obtained the same result, by a
different method and written in a different form (see
Appendix~\ref{app:gausspot}).  Once more, angular momentum
conservation follows straightforwardly:
\begin{align*}
\vec{\tau}_A+\vec{\tau}_B+\vec{r}\times\vec{f} &=
U\mat{\epsilon}:\mat{H}^{-1}\bigl(\mat{A}+\mat{B}\bigr)
+(\vec{r}+\vec{q}_{A}-\vec{q}_{B})\times\vec{f}
\\
&=U\mat{\epsilon}:\bigl(\mat{1}-\ell^2\mat{H}^{-1}\bigr)
+\ell^2\vec{\kappa}\times\vec{f} \;,
\end{align*}
using eqn~\eqref{eqn:newrcscaled}.  The tensor $\mat{1}-\ell^2\mat{H}^{-1}$ is
symmetric, so the double contraction with $\mat{\epsilon}$ gives zero, and
$\vec{f}=U\vec{\kappa}$, so the last cross product also gives zero.

\section{Ellipsoid Contact Potential}
\label{sec:ecp}
As mentioned in section \ref{sec:shifted}, the distance function
appearing in the Gay-Berne and related potentials does not correctly
represent the geometry of ellipsoidal particles.  Formulae reflecting
the true ellipsoid geometry were set out by Perram et al.\
\cite{perram.jw:1996.a}. The distance function takes the more general
form
\begin{equation}
\varphi = \sigma^{-2} = 
\alpha\beta\thinspace
\hat{\vec{r}}\cdot\bigl(\alpha\mat{A}+\beta\mat{B}\bigr)^{-1}\cdot\hat{\vec{r}}
\equiv 
\alpha\beta\thinspace\hat{\vec{r}}\cdot\mat{G}^{-1}\cdot\hat{\vec{r}} \;,
\end{equation}
where $\alpha+\beta=1$.
The matrix
\begin{equation}
\label{eqn:Gdef}
\mat{G} = \alpha\mat{A}+\beta\mat{B}
\end{equation}
plays a similar role to $\mat{H}$ in the previous sections; the case
$\alpha=\beta=\frac{1}{2}$ corresponds to the standard Gay-Berne, or Gaussian
overlap, definition of diameter seen earlier, and in this case
$\mat{G}=\tfrac{1}{2}\mat{H}$. The `true' ellipsoid geometry corresponds to the
case where $\alpha=(1-\beta)$ has been chosen to maximise $\varphi$, i.e.\ 
minimise $\sigma$. This procedure is carried out numerically in the simulation.

The ``scaled potential'' analogous to eqn~\eqref{eqn:Uscaled}
\begin{gather}
U = U\bigl(
r^2/\sigma^2(\hat{\vec{r}},\hat{\mat{a}},\hat{\mat{b}})
\bigr)
 = U(\Phi) \;,
\nonumber \\
\text{where}\quad
\label{eqn:phidefecp}
\Phi = r^2 \varphi = r^2/\sigma^2 = 
\alpha\beta\thinspace
\vec{r}\cdot\bigl(\alpha\mat{A}+\beta\mat{B}\bigr)^{-1}\cdot\vec{r}
\equiv \alpha\beta\thinspace\vec{r}\cdot\mat{G}^{-1}\cdot\vec{r} \;,
\end{gather}
is termed an \emph{ellipsoid contact potential} \cite{perram.jw:1996.a}. The
forces and torques are derived in a manner analogous to that of Section
\ref{sec:scaled}. Define an auxiliary vector $\vec{k}$ by solving the equations
\begin{equation}
\label{eqn:sdef}
\mat{G}\cdot\vec{k} = \vec{r} \qquad \Leftrightarrow \qquad
\vec{k} = \mat{G}^{-1} \cdot \vec{r} \:.
\end{equation}
In terms of $\vec{k}$
\begin{align}
\label{eqn:phidef2ecp}
\Phi &= \alpha\beta \vec{r}\cdot\vec{k} \;,
\\
\frac{\partial \Phi}{\partial\vec{r}} &= 
2\alpha\beta\mat{G}^{-1}\cdot\vec{r}
=
2\alpha\beta\vec{k} \:.
\end{align}
This last formula will give the interparticle force from 
\begin{equation}
\label{eqn:fecp}
\vec{f} 
=
-\frac{\D U}{\D\Phi}
\frac{\partial \Phi}{\partial\vec{r}}
=
-\frac{\D U}{\D\Phi} 2\alpha\beta \vec{k} \:.
\end{equation}
The derivation of torques follows the pattern of previous sections.
The results are
\begin{subequations}
\begin{align}
\vec{\tau}_A &=\frac{\D U}{\D \Phi}2\alpha^2\beta\sum_m 
(\vec{k}\cdot\vec{a}_m)(\vec{a}_{m}\times\vec{k})
=
\frac{\D U}{\D \Phi}2\alpha^2\beta
(\vec{k}\cdot\mat{A}\times\vec{k}) \;,
\\
\vec{\tau}_B &=
\frac{\D U}{\D \Phi}2\alpha\beta^2\sum_m 
(\vec{k}\cdot\vec{b}_m)(\vec{b}_{m}\times\vec{k})
=
\frac{\D U}{\D \Phi}2\alpha\beta^2
(\vec{k}\cdot\mat{B}\times\vec{k}) \:.
\end{align}
\end{subequations}
Once again these may be written
\begin{subequations}
\begin{align}
\vec{\tau}_A 
&= -\alpha\vec{k}\cdot\mat{A}\times\vec{f}
= \vec{r}_{CA}\times\vec{f} \;,
\\
\vec{\tau}_B 
&=-\beta\vec{k}\cdot\mat{B}\times\vec{f}
=\vec{r}_{CB}\times(-\vec{f}) \;,
\end{align}
\end{subequations}
where $\vec{r}_C$ is now defined by
\begin{equation}
\label{eqn:rcecp}
\vec{r}_C = \vec{r}_A - \alpha\mat{A}\cdot\vec{k} 
= \vec{r}_B + \beta\mat{B}\cdot\vec{k}
\end{equation}
(see Appendix~\ref{app:contact}).  Conservation of angular momentum
follows immediately.
One final point \cite{perram.jw:1996.a}: the above derivatives take no account
of the variation of $\alpha, \beta$ as the particles are moved; but these
extra terms vanish because the relevant functions have been minimised/maximised
with respect to variations of these parameters.

The corresponding formulae for more general (for example, shifted) potentials
based on true ellipsoid geometry are
\begin{align}
\varphi &= \alpha\beta\hat{\vec{r}}\cdot\mat{G}^{-1}\cdot\hat{\vec{r}}
= \alpha\beta r^{-1} \hat{\vec{r}}\cdot\vec{k} \;,
\\
\frac{\partial \varphi}{\partial\hat{\vec{r}}} &= 
2\alpha\beta\mat{G}^{-1}\cdot\hat{\vec{r}}
=
2\alpha\beta r^{-1} \vec{k} \:.
\end{align}
The forces and torques are
\begin{subequations}
\begin{align}
\vec{f} 
&=
-\frac{\partial U}{\partial r}\hat{\vec{r}}
-\frac{\partial U}{\partial \varphi}
r^{-1} \frac{\partial \varphi}{\partial\hat{\vec{r}}} \cdot 
\bigl(\mat{1}-\hat{\vec{r}}\otimes\hat{\vec{r}}\bigr)
=
-\frac{\partial U}{\partial r}\hat{\vec{r}}
-\frac{\partial U}{\partial \varphi}
\frac{2\alpha\beta}{r^{2}}
\bigl(\vec{k}-(\vec{k}\cdot\hat{\vec{r}})\hat{\vec{r}}\bigr) \:,
\\
\vec{\tau}_A &=\frac{\partial U}{\partial \varphi}
\frac{2\alpha^2\beta}{r^2}\sum_m 
(\vec{k}\cdot\vec{a}_m)(\vec{a}_{m}\times\vec{k})
=
\frac{\partial U}{\partial \varphi}
\frac{2\alpha^2\beta}{r^2}
(\vec{k}\cdot\mat{A}\times\vec{k}) \;,
\\
\vec{\tau}_B &=
\frac{\partial U}{\partial \varphi}
\frac{2\alpha\beta^2}{r^2}\sum_m
(\vec{k}\cdot\vec{b}_m)(\vec{b}_{m}\times\vec{k})
=
\frac{\partial U}{\partial \varphi}
\frac{2\alpha\beta^2}{r^2}
(\vec{k}\cdot\mat{B}\times\vec{k}) \:.
\end{align}
\end{subequations}
Identifying part of the force 
$\vec{f}_k = -(\partial U/\partial \varphi)(2\alpha\beta/r^2)\vec{k}$
the torques become
\begin{subequations}
\begin{align}
\vec{\tau}_A 
&= -\alpha\vec{k}\cdot\mat{A}\times\vec{f}_k
= \vec{r}_{CA}\times\vec{f}_k \;,
\\
\vec{\tau}_B 
&=-\beta\vec{k}\cdot\mat{B}\times\vec{f}_k
=\vec{r}_{CB}\times(-\vec{f}_k) \;,
\end{align}
\end{subequations}
and, since $\vec{f}-\vec{f}_k$ is parallel to $\hat{\vec{r}}$, angular momentum
conservation follows as before.

A modification of the ellipsoid contact potential, based on the surface-surface
distance calculated along a vector parallel to the line of centres, has recently
been proposed \cite{paramonov.l:2005.a}. Analytical formulae for the forces and
torques are given in that paper.

\section{Conclusions}
In the preceding sections it has been shown how compact expressions
for forces and torques may be derived for pairwise intermolecular
potentials which are written as functions of the molecular orientation
matrices and the centre-centre vector. Illustrations have been given
for various members of the ``extended Gay-Berne family''; in
particular, for the Gaussian potential of Smith and Singer
\cite{smith.w:1985.a}, much simpler expressions for the torques may be
obtained in this way.

The above expressions for forces and torques may be used in molecular
dynamics simulations, as well as hybrid Monte Carlo simulations or
other applications.  They may be straightforwardly combined with
symplectic and reversible integration algorithms
\cite{leimkuhler.bj:2004.a} based either on quaternion parameter
representations of the rigid-body orientations, or on the rotation
matrix itself
\cite{dullweber.a:1997.b,miller.tf:2002.a,kamberaj.h:2005.a}.
\section*{Acknowledgments}
This work was supported by the Engineering and Physical Sciences Research
Council and by the Alexander von Humboldt Foundation.
%
\appendix
\section{The Contact Function}
\label{app:contact}
This appendix recalls the relations leading to the definition of the ``contact
point'' of eqn~\eqref{eqn:rcecp}, and the link with the Berne-Pechukas distance
parameter.  The following function has a unique minimum as $\vec{r}$ varies for
fixed $\alpha$ and $\beta=1-\alpha$ \cite{perram.jw:1984.a,perram.jw:1985.a}:
\begin{equation}
\label{eqn:Psi1}
\Psi(\vec{r};\alpha,\beta) = 
\beta (\vec{r}-\vec{r}_A)\cdot\mat{A}^{-1}\cdot(\vec{r}-\vec{r}_A)
+
\alpha (\vec{r}-\vec{r}_B)\cdot\mat{B}^{-1}\cdot(\vec{r}-\vec{r}_B)
\:.
\end{equation}
The location of this minimum, $\vec{r}=\vec{r}_C$, satisfies 
\begin{equation}
\label{eqn:Psimin}
\frac{1}{2}\left.
\frac{\partial\Psi}{\partial\vec{r}}
\right|_{\vec{r}=\vec{r}_C} 
= 
\beta \mat{A}^{-1}\cdot(\vec{r}_C-\vec{r}_A)
+
\alpha \mat{B}^{-1}\cdot(\vec{r}_C-\vec{r}_B)
= \vec{0} \:.
\end{equation}
The definition $\mat{G}\cdot\vec{k}=\vec{r}_{AB}=\vec{r}$, eqn~\eqref{eqn:sdef},
with $\mat{G}=\alpha\mat{A}+\beta\mat{B}$, eqn~\eqref{eqn:Gdef},
leads to an expression for
$\vec{r}_C$
\begin{align*}
\vec{r} = \mat{G}\cdot\vec{k} 
&= \alpha\mat{A}\cdot\vec{k} + \beta\mat{B}\cdot\vec{k}
= (\vec{r}_A - \vec{r}_C) + (\vec{r}_C - \vec{r}_B) \;,
 \\
\text{where}\quad\vec{r}_C &= \vec{r}_A - \alpha\mat{A}\cdot\vec{k} 
= \vec{r}_B + \beta\mat{B}\cdot\vec{k} \;,
\end{align*}
which satisfies eqn~\eqref{eqn:Psimin}. The value at the minimum is
\begin{align*}
\left.\Psi\right|_{\vec{r}=\vec{r}_C} 
&= \beta(\vec{r}_C - \vec{r}_A)\cdot(-\alpha\vec{k})
+
\alpha(\vec{r}_C - \vec{r}_B)\cdot(\beta\vec{k})
\\
&= \alpha\beta\vec{r}\cdot\vec{k} = 
\alpha\beta\vec{r}\cdot\mat{G}^{-1}\cdot\vec{r} \equiv
\Phi \;,
\end{align*}
as defined in eqn~\eqref{eqn:phidef2ecp}.  Having identified $\vec{r}_C$ and
$\Phi$, the function $\Psi$ may be re-written
\begin{equation}
\label{eqn:Psi2}
\Psi(\vec{r};\alpha,\beta) = 
(\vec{r}-\vec{r}_C)\cdot
\bigl(\beta\mat{A}^{-1}+\alpha\mat{B}^{-1}\bigr)
\cdot(\vec{r}-\vec{r}_C) + \Phi \;,
\end{equation}
as may be checked by expanding and comparing coefficients.

The corresponding equations for the case $\alpha=\beta=\tfrac{1}{2}$ are
conveniently written with the definitions
$\mat{H}\cdot\vec{\kappa}=\vec{r}_{AB}=\vec{r}$, eqn~\eqref{eqn:zetadef},
$\mat{H}=\mat{A}+\mat{B}$, eqn~\eqref{eqn:Hdef}, 
\begin{align}
\Psi(\vec{r}) &= 
\tfrac{1}{2} (\vec{r}-\vec{r}_A)\cdot\mat{A}^{-1}\cdot(\vec{r}-\vec{r}_A)
+
\tfrac{1}{2}
(\vec{r}-\vec{r}_B)\cdot\mat{B}^{-1}\cdot(\vec{r}-\vec{r}_B) \;,
\label{eqn:Psi3}
\\
\vec{r}
 = \mat{H}\cdot\vec{\kappa}
&= \mat{A}\cdot\vec{\kappa}
+ \mat{B}\cdot\vec{\kappa}
= (\vec{r}_A - \vec{r}_C) + (\vec{r}_C - \vec{r}_B) \;,
\\
\vec{r}_C &= \vec{r}_A - \mat{A}\cdot\vec{\kappa} 
= \vec{r}_B + \mat{B}\cdot\vec{\kappa} \;,
\\
\left.\Psi\right|_{\vec{r}=\vec{r}_C} 
&= \tfrac{1}{2}\vec{r}\cdot\vec{\kappa} = 
\tfrac{1}{2}\vec{r}\cdot\mat{H}^{-1}\cdot\vec{r} \equiv
\Phi \;,
\label{eqn:Psi4Phi}
\\
\Psi(\vec{r}) &= 
\tfrac{1}{2}(\vec{r}-\vec{r}_C)\cdot
\bigl(\mat{A}^{-1}+\mat{B}^{-1}\bigr)
\cdot(\vec{r}-\vec{r}_C) + \Phi
\:.
\label{eqn:Psi4}
\end{align}
The generalised Berne-Pechukas overlap function is computed in terms of the
two-centre Gaussian overlap integral
\begin{align}
I &= \int\D\vec{r}~\rho_A(\vec{r}-\vec{r}_A) \rho_B(\vec{r}-\vec{r}_B) 
\nonumber \\
&= \frac{1}{8\pi^3 \sqrt{|\mat{A}||\mat{B}|}}
\int\D\vec{r}~\exp\bigl(-\Psi(\vec{r})\bigr)
= \frac{\exp(-\Phi)}{\sqrt{8\pi^3}\sqrt{|\mat{A}+\mat{B}|}} \;,
\label{eqn:twocenter}
\end{align}
where the densities $\rho_A$, $\rho_B$, are defined in eqns~\eqref{eqn:rhoABdef}
and the function $\Psi(\vec{r})$ is defined in eqn~\eqref{eqn:Psi3}.  The last
step is carried out either by using Fourier transforms \cite{berardi.r:1995.a}
or by re-writing $\Psi(\vec{r})$ in the form of eqn~\eqref{eqn:Psi4} and
shifting origin to $\vec{r}_C$ \cite{perram.jw:1996.a}; $\Phi$ is defined in
eqn~\eqref{eqn:Psi4Phi}.
In practice, the Gaussian form of eqn~\eqref{eqn:twocenter} is not commonly used
as a potential in simulations; instead, the function $\Phi$ is extracted from
it, used to define a range parameter $\sigma$, through $\varphi=\Phi/r^2$
and $\sigma = 1/\sqrt{\varphi}$, eqns~\eqref{eqn:allsigma} of the text; and this
range parameter is used in a variety of potentials.
\section{The Gaussian Potential}
\label{app:gausspot}
The Gaussian potential of Smith and Singer \cite{smith.w:1985.a} is
computed in terms of the double integral
\begin{equation*}
U = \iint\D\vec{r}_a\D\vec{r}_b~
\rho_A(\vec{r}_a-\vec{r}_A) \rho_B(\vec{r}_b-\vec{r}_B)
u(\vec{r}_a-\vec{r}_b) \;,
\end{equation*}
where $u(\vec{r}_a-\vec{r}_b)$ is defined by eqn~\eqref{eqn:sspot}.  
The two integrals are carried out successively: each has the same two-centre
form as eqn~\eqref{eqn:twocenter}, so they are straightforward.
The result is
\begin{equation*}
U 
 = 
C|\mat{H}|^{-1/2}~\exp\bigl(-\tfrac{1}{2}\vec{r}
\cdot\mat{H}^{-1}\cdot\vec{r}\bigr) \;,
\quad \text{where $\vec{r}=\vec{r}_A-\vec{r}_B$ 
and $\mat{H}=\ell^2\mat{1}+\mat{A}+\mat{B}$} \;,
\end{equation*}
and this is eqn~\eqref{eqn:gausspot} of the main text.  Smith and
Singer derived an expression for the torque
arising from the Gaussian potential, from first principles, treating
$u(\vec{r}_a-\vec{r}_b)$ as a site-site potential and integrating
$\vec{r}_{ab}\times\vec{f}_{ab}$ over the density distributions.  In
the notation of sec~\ref{sec:gaussian}, their result for the torque on
$A$ is
\begin{align}
\vec{\tau}_A 
&= -\bigl(\ell^2\mat{1}+\mat{B}\bigr)^{-1} \ast \bigl(
\vec{q}_{A}\otimes\vec{r} + \vec{q}_{A}\otimes\vec{q}_{A} +
\mat{A}-\mat{A}\mat{H}^{-1}\mat{A}
\bigr) U 
\nonumber \\
&= -\bigl(\mat{H}-\mat{A}\bigr)^{-1} \ast \left[ 
\vec{q}_{A}\otimes(\mat{1}-\mat{A}\mat{H}^{-1})\cdot\vec{r} + 
(\mat{1}-\mat{A}\mat{H}^{-1})\mat{A}
\right] U \;,
\label{eqn:sstorque}
\end{align}
where the definition
$\vec{q}_{A}=-\mat{A}\cdot\vec{\kappa}=-\mat{A}\mat{H}^{-1}\cdot\vec{r}$
is used to simplify.  Smith and Singer define
the operation $\ast$ to give a vector from two matrices
$(\mat{A}\ast\mat{B})_k=\epsilon_{ijk}A_{il}B_{jl}$ (note the
conventional summation over all three repeated indices $i$, $j$ and
$l$). This may be recognised as a double contraction of the
Levi-Civita tensor $\mat{\epsilon}$ with a matrix product:
\begin{equation*}
\mat{A}\ast\mat{B} =
-\mat{\epsilon}:\mat{A}\mat{B}^\mathsf{T} =
\mat{\epsilon}:\mat{B}\mat{A}^\mathsf{T} \:.
\end{equation*}
Using this, and noting that
\begin{equation*}
(\mat{H}-\mat{A})^{-1}(\mat{1}-\mat{A}\mat{H}^{-1}) = 
(\mat{H}-\mat{A})^{-1}(\mat{H}-\mat{A})\mat{H}^{-1} = \mat{H}^{-1} \;,
\end{equation*}
eqn~\eqref{eqn:sstorque} simplifies considerably:
\begin{align*}
\vec{\tau}_A 
&= \mat{\epsilon}:\bigl(
\mat{H}^{-1}\cdot\vec{r}\otimes\vec{q}_{A} + \mat{H}^{-1}\mat{A}
\bigr)U 
= \mat{\epsilon}:\bigl(
\vec{\kappa}\otimes\vec{q}_{A} + \mat{H}^{-1}\mat{A}
\bigr)U 
\\
&= \bigl(
\vec{q}_{A}\times\vec{\kappa} + \mat{\epsilon}:\mat{H}^{-1}\mat{A}
\bigr)U 
\:.
\end{align*}
This is eqn~\eqref{eqn:tAzoid} of the main text.

In the derivation of eqn~\eqref{eqn:mpa1}, it is necessary to obtain
the derivative of $|\mat{H}|$ with respect to a component of one of the rotation
matrices. Use
\begin{equation*}
|\mat{H}|' = 
\begin{vmatrix} 
H_{xx}' & H_{xy}' & H_{xz}' \\ 
H_{yx} & H_{yy} & H_{yz} \\ 
H_{zx} & H_{zy} & H_{zz} 
\end{vmatrix}
+
\begin{vmatrix} 
H_{xx} & H_{xy} & H_{xz} \\ 
H_{yx}' & H_{yy}' & H_{yz}' \\ 
H_{zx} & H_{zy} & H_{zz} 
\end{vmatrix}
+
\begin{vmatrix} 
H_{xx} & H_{xy} & H_{xz} \\ 
H_{yx} & H_{yy} & H_{yz} \\ 
H_{zx}' & H_{zy}' & H_{zz}' 
\end{vmatrix} \;.
\end{equation*}
Here, $H_{\mu\nu} = \ell^2\delta_{\mu\nu} + \sum_m a_{m\mu} a_{m\nu} + \sum_n
b_{n\mu} b_{n\nu}$, so $\partial H_{\mu\nu}/\partial a_{m x} = \delta_{\mu
  x}a_{m\nu}+\delta_{\nu x} a_{m\mu}$.
Hence
\begin{align*}
\frac{\partial|\mat{H}|}{\partial a_{m x}} 
&= 
\begin{vmatrix} 
2a_{m x} & a_{m y} & a_{m z} \\ 
H_{yx} & H_{yy} & H_{yz} \\ 
H_{zx} & H_{zy} & H_{zz} 
\end{vmatrix}
+
\begin{vmatrix} 
H_{xx} & H_{xy} & H_{xz} \\ 
a_{m y} & 0 & 0 \\ 
H_{zx} & H_{zy} & H_{zz} 
\end{vmatrix}
+
\begin{vmatrix} 
H_{xx} & H_{xy} & H_{xz} \\ 
H_{yx} & H_{yy} & H_{yz} \\ 
a_{m z} & 0 & 0
\end{vmatrix}
\\
&= 
2a_{m x} 
\begin{vmatrix} 
H_{yy} & H_{yz} \\ 
H_{zy} & H_{zz} 
\end{vmatrix}
-
2a_{m y}
\begin{vmatrix} 
H_{xy} & H_{xz} \\ 
H_{zy} & H_{zz} 
\end{vmatrix}
+
2a_{m z}
\begin{vmatrix} 
H_{xy} & H_{xz} \\ 
H_{yy} & H_{yz} \\ 
\end{vmatrix}
\;,
\end{align*}
using the fact that $\mat{H}$ is symmetric.  Moreover, each (signed) $2\times2$
determinant here is an element of the (signed) cofactor matrix, the transpose of
which is $\mat{H}^{-1}$ multiplied by $|\mat{H}|$.  This allows the desired
derivative to be written
\begin{equation*}
\frac{\partial|\mat{H}|}{\partial a_{m x}} 
=
2|\mat{H}| \left(
a_{m x} \bigl[\mat{H}^{-1}\bigr]_{xx} +
a_{m y} \bigl[\mat{H}^{-1}\bigr]_{yx} +
a_{m z} \bigl[\mat{H}^{-1}\bigr]_{zx}
\right)
\:.
\end{equation*}
Collecting together results for all the components gives
\begin{equation*}
\frac{1}{2}|\mat{H}|^{-1}\frac{\partial|\mat{H}|}{\partial \vec{a}_{m}} 
= \vec{a}_{m}\cdot\mat{H}^{-1} = \mat{H}^{-1}\cdot\vec{a}_{m} \:.
\end{equation*}
This is used in the derivation of eqn~\eqref{eqn:mpa1}.
\newpage
\providecommand{\acronym}[1]{\textsc{\MakeLowercase{#1}}}

\end{document}